\documentclass[aps,prd,12point,twocolumn,nofootinbib,showpacs,superscriptaddress]{revtex4-1}
\def\theequation{\arabic{section}.\arabic{equation}}
\usepackage{psfrag}
\usepackage{subfigure}
\usepackage{color}
\usepackage{mathrsfs}
\usepackage{graphicx}
\usepackage{amssymb, bm}
\usepackage{amsmath, amsthm}
\usepackage{epstopdf}
\usepackage{hyperref}
\usepackage{enumerate}
\usepackage{longtable}
\usepackage{amsmath}
\usepackage{soul}

\newcommand{\be}{\begin{equation}}
\newcommand{\ee}{\end{equation}}

\begin{document}
\def\theequation{\arabic{section}.\arabic{equation}} 

\title{Friedmann-Lema\^itre-Robertson-Walker cosmology through the lens 
of gravitoelectromagnetism}



\author{Valerio Faraoni}
\email[vfaraoni@ubishops.ca]{}
\affiliation{Department of Physics \& Astronomy, Bishop's University, 2600 
College Street, Sherbrooke, Qu\'ebec, Canada J1M~1Z7
}

\author{Genevi\`eve Vachon}
\email[gvachon18@ubishops.ca]{}
\affiliation{Department of Physics \& Astronomy, Bishop's University, 2600 
College Street, Sherbrooke, Qu\'ebec, Canada J1M~1Z7
}

\author{Robert Vanderwee}
\email[rvanderwee20@ubishops.ca]{}
\affiliation{Department of Physics \& Astronomy, Bishop's University, 2600 
College Street, Sherbrooke, Qu\'ebec, Canada J1M~1Z7
}

\author{Sonia Jose}
\email[sjose21@ubishops.ca]{}
\affiliation{Department of Physics \& Astronomy, Bishop's University, 2600 
College Street, Sherbrooke, Qu\'ebec, Canada J1M~1Z7
}


\date{\today}

\begin{abstract}

Friedmann-Lema\^itre-Robertson-Walker cosmology is examined from the point 
of view of gravitoelectromagnetism, in the approximation of spacetime 
regions small in comparison with the Hubble radius. The usual Lorentz 
gauge is not appropriate for this situation, while the 
Painlev\'e-Gullstrand gauge is rather natural. Several non-trivial 
features and differences with respect to ``standard'' asymptotically flat 
gravitoelectromagnetism are discussed.

\end{abstract}


\maketitle

\section{Introduction}
\label{sec:1}
\setcounter{equation}{0}

The weak-field, slow-motion limit of General Relativity (GR) produces 
Newtonian gravity while, by allowing for relativistic motions (but keeping 
the gravitational field weak), one obtains the linearized version of GR. 
It is well known ({\em e.g.}, \cite{Waldbook}) that linearized gravity can 
be recast formally as a Maxwell-like theory by introducing a 
gravitoelectric 
and a gravitomagnetic potential.  Gravitoelectromagnetism has a long 
history and several applications ({\em e.g.}, \cite{Matte53, Braginsky77, 
Teyssandier77, Teyssandier78, Jantzen92, CiufoliniWheeler, 
BonillaSenovilla97, Maartens:1997fg, ClarkTucker00, Senovilla00, 
Ruggiero:2002hz, Mashhoon:2003ax, Ruggiero:2020oxo, Ruggiero:2021uag, 
Ruggiero:2021lpf} and references therein) and it is universally recognized 
as a characteristic of GR. Certain geometries that are solutions of the 
Einstein equations are usually not contemplated from the point of view of 
gravitoelectromagnetism in their weak-field limit.  Here we address the 
gravitoelectromagnetic limit of Friedmann-Lema\^itre-Robertson-Walker 
(FLRW) 
cosmology. The weak-field limit is obtained in the approximation of small 
regions of space around an observer's worldine and small intervals of time 
centered around a particular time (for example, the present time of that 
observer). Comoving observers are commonly used in cosmology, but we will 
introduce also the point of view of radial freely falling observers and 
of Painlev\'e-Gullstrand observers of a de Sitter space osculating the 
FLRW universe. The view of FLRW cosmology through the lens of 
gravitoelectromagnetism is quite unconventional and exhibits several 
differences with respect to ``standard'' linearized GR in asymptotically 
flat spacetimes. In particular, in spite of certain similarities, 
cosmological gravitoelectromagnetism offers the chance to discuss gauges 
different from the usual Lorentz gauge, which are necessarily encountered 
in this context. As expected, because of spatial isotropy the 
gravitomagnetic field vanishes identically, while the 
gravitoelectric field is purely radial. Overall, the contexts of standard 
linearized GR and of the local approximation of FLRW cosmology with a de 
Sitter space are quite different.

To recap, there are three motivations for this work. First, there 
is the curiosity to explore the paradigm of gravitoelectromagnetism in 
cosmology, a context in which (to the best of our knowledge) it has not 
been discussed thus far. Second, we are interested in finding physically 
meaningful 
contexts in which the usual Lorentz gauge does not apply and one needs to 
expand the box of existing tools in gravitoelectromagnetism (the only 
other gauge used in the literature is the Bakolopous-Kanti one discussed 
in Sec.~\ref{sec:3}). Last but not least, everything we know about 
structure formation in the universe comes from $N$-body simulations in the 
early universe. These simulations are Newtonian in spite of the fact that 
they are performed on a box with side equal to a few times the Hubble 
radius. The reason why this is not a problem and Newtonian simulations 
remain accurate has been discussed 
in~\cite{Chisari:2011iq,Green:2011wc,Faraoni:2015kva}: essentially, 
it boils down to the fact that the peculiar velocities of dark matter 
particles are small compared to the Hubble flow at redshift $z\simeq 100$ 
(when the simulations begin), but this statement is extrapolated from 
calculations in a less than transparent way and depends on 
the gauge adopted \cite{Chisari:2011iq}. In any case, it  
sounds like stating that gravitomagnetic effects are negligible in 
comparison with gravitostatic ones (which are Newtonian), and it seems 
to beg for the point of view of gravitolectromagnetism, which we therefore 
develop here for unperturbed and perturbed FLRW cosmology. The 
gauge-invariant approach of 
\cite{Faraoni:2015kva} to the problem of Newtonian cosmological 
perturbations forming early structures is based on splitting the dynamics 
of dark matter particles into a local (Newtonian) part and a cosmological 
part by introducing the fictitious potential $\Phi=-GM_\mathrm{MSH}/R = 
-Gm/R +  H^2 R^2 /2$, where $M_\mathrm{MSH} $ is the 
Misner-Sharp-Hernandez mass 
contained in a sphere of (physical) radius $R$, $m$ is the mass generating 
the local Newtonian perturbation, and $H$ is the Hubble function. The 
splitting of $\Phi$ comes from a splitting of the 
Misner-Sharp-Hernandez mass \cite{Faraoni:2015kva}. This procedure teases 
out the local dynamics from the cosmological expansion in  
a gauge-invariant way but, although it 
makes sense physically, it was based on guessing $\Phi$ rather than 
deriving it rigorously. Here, applying gravitoelectromagnetism to 
perturbed FLRW universes, we show that $\Phi$ is nothing 
but the gravitostatic potential (while the gravitomagnetic contributions 
are negligible).

We follow the notation and conventions of Ref.~\cite{Waldbook}: the metric 
signature is ${-}{+}{+}{+}$, $G$ is Newton's constant, and units are used 
in which the speed of light $c$ is unity. Round brackets around indices 
denote symmetrization.

\section{Linearized general relativity and gravitoelectromagnetism}
\label{sec:2}
\setcounter{equation}{0}

In linearized GR \cite{Waldbook} it is assumed that an 
asymptotically Cartesian coordinate system exists in which the spacetime 
metric assumes the form
\be
g_{\mu\nu}=\eta_{\mu\nu} + h_{\mu\nu} \,,
\ee
where $\eta_{\mu\nu}$ is the Minkowski metric and the perturbations $ 
h_{\mu\nu}$ are small, $| h_{\mu\nu} | \ll 1$.  The metric perturbations 
are 
supposed to be of order ${\cal O}(\epsilon)$, where $\epsilon$ is a small 
dimensionless parameter and, in linearized theory, the Einstein equations 
are written by discarding terms of order higher than ${\cal O}(\epsilon)$. 
The first order Einstein tensor is \cite{Waldbook}
\be
G_{\mu\nu}^{(1)} = 
- \frac{1}{2}\, \partial^{\alpha} \partial_{\alpha} \bar{h}_{\mu\nu}  
+ \partial^{\alpha}\partial_{( \nu} \bar{h}_{\mu ) \alpha}   
-\frac{1}{2}\, \eta_{\mu\nu} \partial^{\alpha} \partial^{\beta} 
\bar{h}_{\alpha\beta}  \,.\label{Gmunu1}
\ee
It is convenient to use the quantity
\be
\bar{h}_{\mu\nu} \equiv h_{\mu\nu} - \frac{1}{2} \, \eta_{\mu\nu}\,  
{h^{\alpha}}_{\alpha} \,,
\ee
where indices are raised and lowered with the unperturbed tensors 
$\eta^{\alpha\beta}$ and $\eta_{\alpha\beta}$.  The Lorentz gauge
\be
\partial^{\mu} \bar{h}_{\mu\nu}=0  \label{Lorentzgauge}
\ee
is then imposed in order to simplify the first order Einstein equations 
$G_{\mu\nu}^{(1)} =8\pi G T_{\mu\nu}$ to 
\be
\partial^{\alpha} \partial_{\alpha} \bar{h}_{\mu\nu} 
 = - 16 \pi G T_{\mu\nu} \,.\label{linearizedEFE}
\ee
The matter energy-momentum tensor is usually assumed to be of the form 
\be
T_{\mu\nu} = \rho \, u_{\mu} u_{\nu} \label{dust}
\ee
describing a dust with energy density $\rho$ and four-velocity field 
$u^{\mu}$.

Gravitoelectromagnetism is introduced by noting that the linearized 
Einstein equations in the Lorentz gauge~(\ref{Lorentzgauge}) assume the 
form of Maxwell equations and that the geodesic equation resembles the 
equation for the Lorentz force acting on a particle of unit charge 
\cite{Waldbook, Matte53, Braginsky77, Teyssandier77, Teyssandier78, 
Jantzen92, CiufoliniWheeler, BonillaSenovilla97, Maartens:1997fg, 
ClarkTucker00, Senovilla00, Ruggiero:2002hz, 
Mashhoon:2003ax,Ruggiero:2020oxo, Ruggiero:2021uag} (there are, however, 
subtleties in the Lorentz force equation when $\phi_{(g)}$ and 
$\vec{A}_{(g)}$ are time-dependent \cite{Bakopoulos:2014exa}). The line 
element is written as
\begin{eqnarray}
ds^2 &=& -\left( 1-2\phi_{(g)} \right) dt^2 +2\vec{A}_{(g)} \cdot d\vec{x} 
\, dt \nonumber\\
&&\nonumber\\
&\, & + \left( 1+2\phi_{(g)} \right) \delta_{ij} dx^i dx^j 
\,,\label{usual}
\end{eqnarray}
from which one reads off the gravitoelectromagnetic potentials 
$\phi_{(g)}$ and $\vec{A}_{(g)}$ \cite{Matte53, Braginsky77, 
Teyssandier77, Teyssandier78, Jantzen92, CiufoliniWheeler, 
BonillaSenovilla97, Maartens:1997fg, ClarkTucker00, Senovilla00, 
Ruggiero:2002hz, Mashhoon:2003ax,Ruggiero:2020oxo, Ruggiero:2021uag}.

The 3-dimensional projection of the timelike geodesic equation for a 
massive particle of 3-velocity $\vec{v}$ assumes the form analogous to 
the Lorentz force equation \cite{Waldbook}
\be
\vec{a} = -\vec{E}_{(g)} -4\vec{v} \times \vec{B}_{(g)} \,.
\ee
In the following, we develop gravitoelectromagnetism for FLRW cosmology 
and we compare it with the ``standard'' version summarized in this 
section.

\section{Gravitoelectromagnetism in FLRW spacetime}
\label{sec:3}
\setcounter{equation}{0}

Let us consider now the FLRW metric in comoving coordinates $\left( t, x, 
y, z \right)$
\begin{eqnarray}
ds^2 &=& -dt^2 + a^2(t) \left( dx^2 +dy^2 +dz^2 \right)\\
&&\nonumber\\
&=&-dt^2 + a^2(t) \left( dr^2 +r^2 d\Omega_{(2)}^2 \right) \,,\label{FLRW}
\end{eqnarray}
 where the last line uses polar comoving coordinates 
$\left( t, r, \vartheta, \varphi \right)$
 and $d\Omega_{(2)}^2\equiv d\vartheta^2 + \sin^2 \vartheta 
\, 
d\varphi^2$ is the line element on the unit 2-sphere. The areal 
radius is
\be
R(t,r)=a(t) \, r =\sqrt{ X^2 +Y^2 +Z^2} \,,
\ee
where $X^i \equiv a(t) \, x^i$ ($i=1,2,3$) are  
(oriented) physical lengths along the $x^i$ axes, while the comoving 
coordinates $x^i$ instead follow the expansion of the cosmic fluid. More 
precisely, two points located on the $x^i$-axis and separated by the 
comoving infinitesimal distance $dx^i$ have physical separation $a(t) 
dx^i$ at time $t$. Two such points at finite comoving distance $x^i$ have 
physical separation $X^i=a(t) x^i$ (however, $dX^i $ does not coincide 
with the physical infinitesimal separation $a(t) dx^i $ unless $a(t)$ is 
approximated with its value $a(t_0) $ at the time $t_0$). 

In order to write the FLRW metric as a formal Minkowski metric 
plus small perturbations, it is convenient to switch 
to the use of coordinates $X^i$ instead of $x^i$, and of the 
areal 
radius $R$ as the radial  coordinate insted of the comoving $r$. We have
\be
dx^i = \frac{ dX^i -HX^i dt}{a} \,, \quad \quad dr=\frac{dR-HRdt}{a} \,,
\ee
where $H \equiv \dot{a}/a$ is the Hubble function and an overdot denotes 
differentiation with respect to the comoving time $t$. Substituting into 
the FLRW line element~(\ref{FLRW}), one obtains \cite{Faraoni:2020ehi} 
\begin{eqnarray}
ds^2& = & -\left(1-H^2R^2\right) dt^2  -2H X^i dtdX^i  +dX^2 + dY^2  
\nonumber\\
&&\nonumber +dZ^2 \nonumber\\
&&\nonumber\\
&=&  -\left(1-H^2R^2\right) dt^2  -2H R \, dt\, dR  +dR^2 + R^2  
d\Omega_{(2)}^2 \nonumber\\
&&\nonumber\\
&=& \left( \eta_{\mu\nu} +h_{\mu\nu} \right)   dX^{\mu} dX^{\nu}   
\label{linearized} 
\end{eqnarray}
where, in the last line,\footnote{We stress that, in the line 
element~(\ref{linearized}), $t$ is still the comoving time and the only 
difference with respect to Eq.~(\ref{FLRW}) is the coordinate switch $x^i 
\rightarrow X^i$: we are now considering  observers 
using a Schwarzschild-like radius and moving radially with respect to 
the comoving observers.}  the 
metric is formally the Minkowski metric 
$\eta_{\mu\nu}$ plus a deviation $h_{\mu\nu}$ from it that, at this 
stage, is not yet required to be small. Explicitly, we have
\be
h_{00}= H^2 R^2 \,, \quad\quad h_{0i} =-HX^i \,, \quad\quad h_{ij}=0 
\,.\label{corrections}
\ee
This form of the metric resembles linearized gravitational 
theory where the $h_{\mu\nu} $ are small. To establish a parallel with 
linearized GR, we now assume that the 
corrections to the formal 
Minkowski metric appearing in Eq.~(\ref{linearized}) are small. There is 
a conceptual difference with respect to  ``standard'' linearized GR. While 
usually one assumes the existence of an asymptotically 
Cartesian coordinate system in which the metric splits as  
$g_{\mu\nu} = \eta_{\mu\nu} +h_{\mu\nu} $ \cite{Waldbook}, in 
cosmology we have the opposite situation. Spacetime is asymptotically 
(indeed, exactly) FLRW and one obtains $ | h_{\mu\nu}| \ll 1$ only by 
restricting to spacetime  regions small with respect to the Hubble radius 
$H^{-1}$, which implies 
\be
H |X^i | \leq H R \ll 1 \,.
\ee
The physical meaning of this approximation is that spacetime is locally 
flat and the effects of the cosmological 
expansion can only be felt by systems of size non-negligible   
with respect to the radius of curvature of spacetime, in this case the 
Hubble radius 
$H^{-1}$. However,  
Eqs.~(\ref{linearized}) and~(\ref{corrections}) are exact, no expansion is 
required for their validity, and the $h_{\mu\nu}$ are not {\em a priori} 
small. It is  only when one wants the $h_{\mu\nu}$ to be small in order to 
mirror linearized gravity, and to introduce gravitoelectromagnetism 
(which is our goal here), that 
one restricts oneself to regions much 
smaller than $H^{-1}$ and uses $ \epsilon \equiv HR$ as a smallness 
parameter.

In practice, when one studies cosmological physics in the neighborhood of  
a certain instant of time $t_0$, for example the present time in the 
history of the universe in cosmography, one expands the Hubble function 
$H(t)$ around 
the present time $t_0$. If one allows $|t-t_0|$ to be arbitrary, then 
light signals reaching the observer at time $t_0$ can arrive from distant 
regions of the universe, breaking the assumption that only regions with $H 
R\ll 1$ are 
considered. Therefore, as done in cosmography, we replace the Hubble 
function $H(t)$ with its 
value $H_0\equiv H(t_0)$ and we consider only time intervals such that $ 
H_0 
|t-t_0| \ll 1$, in addition to restricting to regions with $H_0 R \ll 1$. 
The local deviations of the spacetime metric  from the 
Minkowski one then read
\be
h_{00}= H_0^2 R^2 \,, \quad\quad h_{0i} =-H_0 X^i \,, \quad\quad h_{ij}=0 
\ee
in coordinates $\left( t, X^i \right)$ or, with equivalent terminology, 
in the gauge in which the line element assumes the form
\begin{eqnarray}
ds^2 &=& \left( \eta_{\mu\nu} +h_{\mu\nu} \right) dX^{\mu} dX^{\nu} 
\nonumber\\
&&\nonumber\\
&=& -\left(1-H_0^2 R^2\right) dt^2  -2H_0 X^i dtdX^i  + \delta_{ij} dX^i 
dX^j  \,.\nonumber\\
&& \label{LE}
\end{eqnarray}
The approximation $H(t)\simeq H_0=$~const. is equivalent to replacing the 
exact FLRW manifold with a de Sitter spacetime with Hubble 
constant equal to the value 
$H_0 \equiv H(t_0)$ of the Hubble function of the real FLRW spacetime.

As in linearized gravity \cite{Waldbook}, one can introduce 
$  \bar{h}_{\mu\nu} \equiv h_{\mu\nu}-\frac{1}{2} \, \eta_{\mu\nu} 
{h^{\alpha}}_{\alpha} $, which has the only non-vanishing components
\begin{eqnarray}
\bar{h}_{00} &=& \frac{H_0^2 R^2}{2} \,,\label{ourgauge1}\\
&&\nonumber\\
\bar{h}_{0i} &=& \bar{h}_{i0} =-H_0 X^i \,, \label{ourgauge2}\\
&&\nonumber\\
\bar{h}_{ij} &=& \frac{H_0^2 R^2}{2}\,  \delta_{ij} \,, \label{ourgauge3}
\end{eqnarray}
in coordinates $\left( t, X^i \right)$, in which the line 
element~(\ref{LE}) can be written as 
\be
ds^2 =-\left( 1-2\Phi \right)dt^2 + 2A_i dX^i dt + \delta_{ij} dX^i dX^j 
\,.\label{GEMmetric}
\ee
Here 
\be
\Phi = \frac{H_0^2 R^2}{2} \,, \quad \quad \vec{A}= -H_0 \, \vec{X} 
\ee
can be regarded as the gravitoelectric and gravitomagnetic potentials, 
respectively.  
There is, however, something very 
unconventional about this identification: usually \cite{Waldbook}, the 
linearization of the Einstein equations and the formulation of 
gravitoelectromagnetism are performed by imposing the Lorentz gauge 
$\partial^{\mu} \bar{h}_{\mu\nu}=0$ 
 in which the linearized Einstein equations simplify and the resulting 
line element assumes the form~({\ref{usual}). Here, instead, the line 
element appears in the different form~(\ref{LE}). Our 
gauge~(\ref{ourgauge1})-(\ref{ourgauge3}) is incompatible with the Lorentz 
gauge because $\partial^{\mu} \bar{h}_{\mu 
0}=-3H_0 \neq 0$. Is this  a problem? {\em A priori}, it isn't: the 
gravitoelectromagnetic potentials are gauge-dependent and 
the gravitoelectric and gravitomagnetic fields are gauge-independent, as 
expected \cite{ClarkTucker00, Bakopoulos:2014exa}. Clearly, 
the metric looks different in the two gauges and physical interpretations 
based on such gauges will be different. 

There is, however, a more substantial conceptual and gauge-independent  
difference between 
standard linearized gravity and the 
linearized version of cosmology. In the former, the matter stress-energy 
tensor $T_{\mu\nu}$ is assumed to describe a dust (Eq.~(\ref{dust})). In 
the cosmological context, instead, we have replaced 
the exact FLRW space with its de Sitter approximation at time $t_0$, which 
means that $T_{\mu\nu}$ has necessarily the form of the 
effective energy-momentum tensor of a cosmological 
constant $\Lambda=3H_0^2$, 
\be
T_{\mu\nu}= -\Lambda \, g_{\mu\nu}= -3H_0^2 \left( 
\eta_{\mu\nu}+h_{\mu\nu} \right) \,.\label{TLambda}
\ee
Contrary to a dust, this effective stress-energy tensor has 
non-vanishing pressure 
\be
P_{\Lambda}=-\rho_{\Lambda}=  -\frac{\Lambda}{8\pi G} = 
-\frac{3H_0^2}{8\pi G} \,;
\ee
it depends in an essential way from the metric 
perturbations $h_{\mu\nu}$. In particular, in the 
gauge~(\ref{ourgauge1})-(\ref{ourgauge3}) adopted, the 
non-diagonal 
components 
\be
T_{0i} = -3H_0^2 h_{0i}  =3H_0^3 X^i 
\ee
describe an energy current which is generated by the transformation 
from the comoving coordinates $x^i$ (which expand with the cosmic 
substratum) to the (oriented) physical lengths $X^i$ along the spatial 
axes. This means that the observers at rest in coordinates $\left( t, X^i 
\right)$ (which we call ``Schwarzschild-like observers'' because they use 
the Schwarzschild-like areal radius as the radial coordinate) move 
radially 
with respect to the comoving observers and see a spatial current of 
radially moving matter, while comoving observers see the cosmic 
fluid at rest. This current is due to the use of spatial coordinates not 
adapted to the 
spatial symmetries. The de Sitter approximation 
to the FLRW metric satisfies the Einstein-Friedmann equations (here listed 
in comoving coordinates) 
\be
H^2=\frac{8\pi G}{3} \, \rho -\frac{K}{a^2}  \,,\label{Friedmann}
\ee
\be
\frac{\ddot{a}}{a} = -\frac{4\pi G}{3} \left( \rho +3P \right) \,,
\ee
\be
\dot{\rho}+3H\left( P+\rho \right) =0 \,,
\ee
with $K=0$ in the approximation 
\be
H(t) \simeq H_0 \,, \quad\quad \rho(t) \simeq \rho(t_0) = 
\frac{3H_0^2}{8\pi G} \simeq - P(t)  
\ee
and 
\be
a(t) =a_0 \, \mbox{e}^{ \int H(t) dt } \simeq \, \mbox{e}^{H_0 t} \,,
\ee
where we set $a_0=1$ for convenience.

There is another important difference between standard linearized gravity 
and the local approximation to cosmology: usually, one assumes that $ 
h_{\mu\nu}={\cal O}(\epsilon )$, where $\epsilon $ is a smallness 
parameter, and keeps only 
terms of order $\epsilon$ in the Einstein equations while discarding 
higher order terms.\footnote{We are not concerned here with expansions in 
inverse powers of the speed of light, which one finds in standard 
linearized GR.} In our expansion of the FLRW 
metric, we have metric 
components with different orders of magnitude in the dimensionless 
expansion parameter $ \epsilon = H_0 R $: 
\be 
h_{00}={\cal O}(\epsilon^2) \,, \quad\quad 
h_{0i}={\cal O}(\epsilon )\,, 
\ee
while the $h_{ij}$ are {\it exactly} zero. As a consequence, our context 
is not the usual first order GR and the comparison of 
results is necessarily limited. In particular, we should not expect a 
one-to-one correspondence between these two contexts. With this {\em 
caveat}, let us proceed.

As expected from the spherical symmetry about every spatial point, the 
gravitoelectric field $\vec{E}_{(g)}=-\vec{\nabla}\Phi $ is purely 
radial. The  gravitomagnetic potential $\vec{A}=-H_0 \, \vec{X}$ is 
also purely radial and the gravitomagnetic 
field then vanishes, 
\be
\vec{B}_{(g)}=\vec{\nabla} \times \vec{A} =0 \, .
\ee
The spatial acceleration of a test particle of unit mass is 
\be
\vec{a}= \vec{E}_{(g)} = - \vec{\nabla}\Phi = - H_0^2  R \, \vec{e}_{R}
\ee
where $\vec{e}_{R} $ is the spatial unit vector in the radial direction 
in coordinates $\left( t, \vec{X} \right)$. Moreover, in the approximation 
made $H(t) \simeq H(t_0) \equiv H_0 $, the gravitoelectric and 
gravitomagnetic potentials are time-independent,
\be
\frac{\partial\Phi}{\partial t} = 
\frac{\partial A_i}{\partial t} = 0\,,
\ee
which removes certain unpleasant terms in the Lorentz force equation  
associated with the time dependence and reported, {\em e.g.}, in 
Ref.~\cite{Bakopoulos:2016rkl}.

\subsection{Bakopoulos-Kanti gauge}

A gauge similar to the one used in this section is reported in linearized 
GR by Bakopolous and 
Kanti~\cite{Bakopoulos:2014exa,Bakopoulos:2016rkl}. This is the 
only instance that we are aware of in which gravitoelectromagnetism is 
discussed in a gauge different from the Lorentz 
gauge~(\ref{Lorentzgauge}). 
Specifically, in the context of the linearized theory summarized in 
Sec.~\ref{sec:2}, the Bakopolous-Kanti gauge is \cite{Bakopoulos:2014exa, 
Bakopoulos:2016rkl}
\be
\bar{h}_{00}=\phi_{(g)} \,, \quad\quad \bar{h}_{0i} =-A_i^{(g)} \,, \quad 
\quad 
\bar{h}_{ij}=\phi_{(g)} \, \delta_{ij} 
\ee
or, equivalently, 
\be
h_{00}= 2\phi \,, \quad\quad h_{0i} =-A_i^{(g)} \,, \quad \quad 
h_{ij}=0  \,;
\ee
these authors derive the result that this gauge choice is only possible 
{\em in vacuo}, $T_{\mu\nu}=0$. At first sight, this result seems to 
conflict with the gauge that we obtained in FLRW space, but this 
conclusion would be incorrect. In fact, the two contexts are 
quite different: 
first, Bakopolous and Kanti \cite{Bakopoulos:2014exa, Bakopoulos:2016rkl} 
assume the stress-energy tensor of a dust, while we assume that of a 
cosmological constant~(\ref{TLambda}). Second, in standard linearized 
theory the metric perturbations are all of the same (first) order 
$h_{\mu\nu}={\cal O}(\epsilon)$, while this is not true in the de Sitter 
space approximating a FLRW universe. Indeed, by denoting loosely with $R$ 
the radius of curvature of spacetime, the standard linearized Einstein 
equations~(\ref{linearizedEFE}) give, in order of magnitude, $\epsilon/R^2 
\sim \rho $, where all terms of order higher than ${\cal O}(\epsilon)$ are 
discarded. In the cosmological case, the stress-energy 
tensor~(\ref{TLambda}) proportional to $H_0$ gives, instead, an equation 
of the form $ h \simeq H_0^2 R^2 ={\cal O}(\epsilon^2)$, where the right 
hand side is {\it of second order} in the smallness parameter 
$\epsilon=H_0 R$. Therefore, this right hand side would be dropped from 
the 
linearized field equations in ``standard'' theory and one would conclude 
that this gauge only applies to vacuum, but the cosmological context is 
quite different from the usual linearized theory (moreover, vacuum 
cosmology without $\Lambda$ is meaningless). The procedure that we 
followed, and the standard results on Painlev\'e-Gullstrand coordinates 
for static spherical spacetimes that we discuss in the next section and 
that agree with the previous procedure, are legitimate and do not 
contradict Ref.~\cite{Bakopoulos:2014exa} because of the different 
assumptions.

In the light of the fact that FLRW spacetimes are spherically 
symmetric, we can think of the Bakopolous-Kanti gauge in such situations. 
By virtue of the Jebsen-Birkhoff theorem \cite{Waldbook}, if a linearized  
geometry is expressed 
in the Bakopolous-Kanti gauge {\em and} is spherical, it must be 
the linearization of the  Schwarzschild spacetime 
\be
ds^2 =-\left( 1-\frac{2Gm}{r} \right) dt^2 + \frac{dr^2}{1-2Gm/r} +r^2 
d\Omega_{(2)}^2 
\ee
because it is a vacuum, spherical, and 
asymptotically flat solution of the Einstein equations (this conclusion 
applies also to the spacetime outside spherical black holes in most 		
scalar-tensor theories of gravity in ``reasonable'' situations, see 
\cite{HawkingSTBH, Bekenstein:1996pn, Sotiriou:2011dz, 
Bhattacharya:2015iha}). Indeed, it is not 
even necessary to linearize the Schwarzschild metric to recast it in the 
Painlev\'e-Gullstrand gauge \cite{Painleve, Gullstrand, Martel:2000rn}
\be
ds^2 =-\left( 1-\frac{2Gm}{r} \right) dT^2 + 2 \sqrt{ \frac{2Gm}{r}} \, 
dT \, dr +dr^2 +r^2 d\Omega_{(2)}^2 
\ee
where 
\be
T= t + 4m\left( \sqrt{ \frac{r}{2Gm} } \, + \frac{1}{2} \ln \left| \frac{ 
\sqrt{ \frac{r}{2Gm} }  -1 }{
\sqrt{ \frac{r}{2Gm} } +1 } \right| \right)
\ee
is the Painlev\'e-Gullstrand time \cite{Painleve, Gullstrand, 
Martel:2000rn}. This gauge coincides with the Bakopolous-Kanti gauge 
without the need to assume $ | h_{\mu\nu} | \ll 1$. This situation is 
well-known and we conclude that the Bakopolous-Kanti gauge is most 
interesting in non-spherical situations.

\section{Relation with Painlev\'e-Gullstrand observers}
\label{sec:}
\setcounter{equation}{0}

The Schwarzschild-like observers used in the previous section to discuss 
gravitoelectromagnetism in FLRW cosmology employ the comoving time $t$ but 
differ from comoving observers, with respect to which they move radially.  
In the first part of this section we recall known material from a variety 
of sources  in the literature with the purpose of elucidating the physical 
meaning of these observers (which we do at the end of this section). 

First, let us recall the transformation from comoving to 
Schwarzschild-like coordinates for spatially flat FLRW universes and, in 
particular, for the special de Sitter case that we use to approximate a 
FLRW universe. Beginning from the spatially flat FLRW metric in comoving 
coordinates~(\ref{FLRW}) and using the areal radius  $ R\equiv a(t)r $,  
we have obtained the line element
\be
ds^2 = -\left(1-H^2 R^2\right) dt^2  -2H X^i dtdX^i  + \delta_{ij} 
dX^i dX^j  \,.
\ee
The cross-term in $dt dR$ can now be eliminated by introducing the new 
time $T$ defined by
\be
dT=\frac{1}{F} \left( dt+ \beta dR \right) \,, \label{dT}
\ee
where $ F ( t, R)  $ is an integrating factor 
satisfying
\be
\frac{\partial}{\partial R}\left( \frac{1}{F} \right)=
\frac{\partial}{\partial t}\left( \frac{ \beta}{F} \right)
\label{eqF}
\ee
to guarantee that $dT$ is a locally exact differential, while  
$\beta  \left( t, r \right) $ is, for the moment, an unknown function 
\cite{Faraoni:2015ula}. Substituting  $  dt=FdT-\beta dR $ into the line 
element yields 
\begin{eqnarray}
ds^2  & = & - \left( 1-H^2R^2 \right)  F^2 dT^2 \nonumber\\
&&\nonumber\\
&\, & +2F \left[ \left(1-H^2R^2\right)\beta  -HR \right] dTdR 
\nonumber\\
&&\nonumber\\
& \, & + \left[ 1- \left(1-H^2R^2 \right) \beta^2 +2\beta HR 
\right] dR^2 + R^2 d\Omega_{(2)}^2 \,.\nonumber\\
&& 
\end{eqnarray}
Setting 
\be
\beta \left( t, R \right) = \frac{HR}{1-H^2R^2}  \label{beta}
\ee
reduces the FLRW line element to its Schwarzschild-like form 
\be \label{k=0Nolangauge}
ds^2=-\left( 1-H^2R^2 \right)F^2 dT^2 + 
\frac{dR^2}{1-H^2R^2}+R^2 
d\Omega_{(2)}^2 \,.
\ee
In the special case of de Sitter space the Hubble function $H$ is 
constant and $F=1$ satisfies Eq.~(\ref{eqF}), which 
transforms~(\ref{k=0Nolangauge}) into the de Sitter line element in static 
coordinates. As done in the previous section, we approximate the 
spatially flat FLRW  space with a de Sitter space by replacing  
$H(t)$ with $H_0 \equiv H(t_0)$ around a fixed time $t_0$. The result is 
\be \label{dSstatic}
ds^2 \simeq  - \left( 1-H_0^2 R^2 \right) dT^2 
+\frac{dR^2}{1-H_0^2 R^2} +R^2 d\Omega_{(2)}^2   
\ee
for $H_0 R\ll 1$.

Let us review now the Painlev\'e-Gullstrand coordinates for de Sitter 
space, which are a special case of the more general Martel-Poisson family 
\cite{Martel:2000rn} derived in \cite{Faraoni:2020ehi} for de Sitter 
space.

Begin from the de Sitter line element in Schwarzschild-like 
coordinates and define a new time coordinate $\bar{T}$ 
by 
\be
d\bar{T}= dT + \frac{ \sqrt{1-pf}}{f} \, dR \,, \label{dSdT}
\ee
where $f \equiv 1-H_0^2 R^2$ and $p$ is a parameter labelling different 
charts (it is straightforward to check that the differential  
$d\bar{T} $ is exact). The physical meaning of $p$ is obtained by writing  
the equation of outgoing ($\dot{R}>0$) radial timelike geodesics  
\cite{Faraoni:2020ehi,Vachon:2021bya} 
\be
\frac{ds^2}{d\tau^2} = -f \left( \frac{dT}{d\tau} \right)^2 +\frac{1}{f} 
\, \left( \frac{dR}{d\tau} \right)^2 =-1 \,,\label{dStimelikeradial1}
\ee
where $\tau$ is the proper time along timelike geodesics. Because of 
the presence of the timelike Killing vector $T^a= \left( \partial/\partial 
T \right)^a $  in the  de 
Sitter metric approximating the FLRW universe, the 
energy is conserved along these radial timelike geodesics and, denoting 
with $p^{c}=m u^c$ the four-momentum of a particle of mass $m$ 
and four-velocity $u^c$, $
p_a T^a= -E $  is constant  along the geodesic. If $\bar{E} 
\equiv E/m$ denotes the particle energy per 
unit mass, then $ 
u^0= dT/d\tau = \bar{E}/f$, 
\be
\left( \frac{dR}{d\tau} \right)^2= \bar{E}^2-f \,,
\ee
and 
\be
\frac{dR}{d\tau} = \pm \sqrt{ \bar{E}^2-f} \,,\label{laminch}
\ee
where the upper sign refers to outgoing and the lower sign to ingoing 
geodesics.  Introducing $p\equiv 1/\bar{E}^2 $, the radial component of 
the four-velocity reads \cite{Faraoni:2020ehi}
\be
\frac{dR}{d\tau}=\frac{dR}{dt}\, \frac{dt}{d\tau} =\pm \gamma(v) \, v 
=\pm \frac{v}{\sqrt{1-v^2}} =\pm \sqrt{\bar{E}^2-f} \,,
\ee
where $\gamma(v)$ is the Lorentz factor and $v=|\vec{v}|$ is the 
magnitude of the coordinate 3-velocity. 

At the origin it is  
\begin{eqnarray}
&& \left| \frac{dR}{d\tau}\Big|_{R=0} \right| = 
\frac{v_0}{\sqrt{1-v_0^2}}=
\sqrt{\bar{E}^2-1} \,,\\
&&\nonumber\\
&& p\equiv \frac{1}{\bar{E}^2} = {1-v_0^2} \,,
\end{eqnarray}
and the parameter $p$ spans the range  $ 0 < p \leq 1 $ (this is similar 
to the case of Martel-Poisson coordinates in Schwarschild space 
\cite{Martel:2000rn}). 

The outgoing ``Martel-Poisson'' observer freely-falling from rest from 
the origin $R=0$ perceives the geometry 
\be
ds^2=-fd\bar{T}^2 + 2 \sqrt{1-pf} \,d\bar{T} dR +p dR^2 +R^2 
d\Omega_{(2)}^2 \label{dSlinewithp}
\ee
where the time coordinate $\bar{T}$ is given explicitly by 
\cite{Faraoni:2020ehi} 
\begin{eqnarray}
\bar{T} &=& T + \sqrt{1-p} \int 
dR \, \frac{ \sqrt{1+\frac{p}{1-p} \, H^2_0 R^2}}{ 
1-H_0^2R^2} \nonumber\\
&&\nonumber\\
&=& T + \frac{\sqrt{p} }{H_0} \left[ 
\frac{1}{p}  \tanh^{-1} \left( \frac{1}{ \sqrt{p(1-p)}} \, 
\frac{H_0 R  }{ \sqrt{1+  \frac{p}{1-p} \, H_0^2 R^2} } \right) 
\right. \nonumber\\
&&\nonumber\\
&\, & \left. -\sinh^{-1} \left( 
\sqrt{ \frac{p}{1-p}}\,  H_0 R \right)  \right] + \mbox{const.}
\end{eqnarray}

The special parameter value $p=1$ gives Painlev\'e-Gullstrand coordinates 
(see \cite{Vachon:2021bya} for a discussion of different radial geodesic 
observers in FLRW 
cosmology) and it is now clear that it corresponds to vanishing initial 
velocity $v_0=0$ of the freely-falling observer at the origin. With 
$p=1$, the de Sitter  line element~(\ref{dSlinewithp}) assumes the 
Painlev\'e-Gullstrand form  \cite{Faraoni:2020ehi}
\be
ds^2=-fd\bar{T}^2 + 2H_0 R \,d\bar{T} dR + dR^2 +R^2 
d\Omega_{(2)}^2  \,.\label{PGform}
\ee
The time slices are flat and the 
Painlev\'e-Gullstrand time is simply \cite{Faraoni:2020ehi}
\begin{eqnarray}
\bar{T} &=& T - \frac{1}{2H_0 } \ln 
\left|1-H_0^2 R^2\right| +\mbox{const.},  \label{Tbar} 
\end{eqnarray}
which was used in previous literature~\cite{Parikh:2002qh}.

The Schwarzschild-like observers seeing the geometry~(\ref{LE}) and using 
comoving time $t$ and coordinates $X^i=a(t) \, x^i$ are not 
Painlev\'e-Gullstrand observers, although the line element~(\ref{LE}) has 
the Painlev\'e-Gullstrand form with flat spatial sections. The reason is 
that all freely-falling observers are related by a Lorentz boost and do 
not accelerate with respect to each other (indeed, in a general spacetime 
freely-falling observers, which do not accelerate with respect to each 
other, are determined up to a Lorentz transformation 
\cite{Weinberg}). The line element~(\ref{LE}) is Lorentz-invariant and has 
the same form for all these observers boosted with respect to 
Painlev\'e-Gullstrand ones. However, the special initial condition $v_0=0$ 
at $R=0$ is satisfied only by Painlev\'e-Gullstrand observers (using the 
time $\bar{T}$) and not by all those Lorentz-boosted with respect to them.

\subsection{Geodesic  observers in FLRW and de Sitter}

In a FLRW universe sourced by a perfect fluid, the comoving observers are 
not, in general, geodesic because they are subject to the pressure 
gradient $\nabla^{\mu} P$ and they accelerate. Because of spatial 
isotropy, $P=P(t)$ and $\nabla^{\mu} P $ points in the direction of 
comoving time. In de Sitter space the pressure $P = -\frac{\Lambda}{8\pi 
G}$ is constant, $\nabla^{\mu} P$ vanishes 
identically and the comoving 
observers of the effective fluid in 
de Sitter space are geodesic. Therefore, freely-falling and comoving 
observers in de Sitter space differ only by a Lorentz boost, which 
agrees with what we have already found with different considerations. 
Painlev\'e-Gullstrand observers are special radial geodesic observers, as 
shown above.

\subsection{FLRW gravitoelectromagnetism and quasilocal mass}

It is well-known \cite{Martel:2000rn,Abreu:2010ru, Nielsen:2005af} that 
the line element of a spherically symmetric (possibly time-dependent) 
spacetime can be recast in the Painlev\'e-Gullstrand form
\begin{eqnarray}
ds^2 &=& -\left( 1-\frac{2GM_\mathrm{MSH}( \bar{t},R)}{R} \right) d\bar{t}^2
\nonumber\\
&&\nonumber\\
&\, & \pm 2 \sqrt{ \frac{2GM_\mathrm{MSH}( \bar{t},R)}{R} } \, d\bar{t} \, 
dR + dR^2  + R^2 d\Omega_{(2)}^2 \,,\nonumber\\
&&   \label{AbreuVisser} 
\end{eqnarray}
where $R$ is the areal radius, $M_\mathrm{MSH} \left( \bar{t}, R 
\right)$ is  the Misner-Sharp-Hernandez mass of  a sphere of radius $R$, 
and one can choose either sign in front of the time-radius cross-term 
(see the discussion in \cite{Faraoni:2020ehi}). The 
expression~(\ref{AbreuVisser}) holds when $M_\mathrm{MSH}$ is 
non-negative. The 
Misner-Sharp-Hernandez mass is defined by\footnote{Since the areal radius 
$R$ is defined only in spherical symmetry, so is the 
Misner-Sharp-Hernandez mass \cite{MSH1,MSH2}.} \cite{MSH1,MSH2}
\be
1-\frac{ 2G M_\mathrm{MSH}}{R} \equiv \nabla^c R \nabla_c R \label{MSH} 
\,.
\ee
This definition is 
expressed by a scalar equation, therefore $M_\mathrm{MSH}$ is 
coordinate-invariant. The Hawking-Hayward quasilocal mass \cite{Hawking, 
Hayward:1993ph} reduces to the Misner-Sharp-Hernandez mass in spherical 
symmetry \cite{Hayward:1994bu} and, in this case, it is the Noether charge 
associated with the conservation of the Kodama current and with spherical 
symmetry  
\cite{Hayward:1994bu}. In general, however, Painlev\'e-Gullstrand 
observers with zero initial velocity cannot be used in non-static 
(spherical) spacetimes because their introduction makes use of energy 
conservation along 
radial timelike geodesics \cite{Martel:2000rn,Faraoni:2020ehi}. Before 
approximating $H(t)$ with $H(t_0)$ in the spatially flat FLRW universe, 
one can introduce the coordinates $\left( t, X^i \right)$ which turn the 
FLRW line element into what looks like the Painlev\'e-Gullstrand form  
with flat spatial sections. However, these coordinates are not those 
associated with freely-falling radial observers with zero initial velocity 
until the approximation $H(t) \simeq H_0$ is made:  
Painlev\'e-Gullstrand observers can be introduced in the Sitter space, but 
not in general (non-static) FLRW universes \cite{Faraoni:2020ehi}.

As noted, Painlev\'e-Gullstrand coordinates are not defined in spherical 
spacetimes or spacetime regions in which the Misner-Sharp-Hernandez mass 
becomes negative. This is the case, {\em e.g.}, of anti-de Sitter space 
with the physical interpretation that the repulsion of the negative 
cosmological constant prohibits a freely-falling observer with zero 
initial velocity from leaving the origin $R=0$ \cite{Faraoni:2020ehi}. 
When $M \geq 0$ and Painlev\'e-Gullstrand coordinates are defined, their 
characterizing feature is that spatial sections are flat.

In a spatially flat FLRW universe, the Misner-Sharp-Hernandez mass defined 
by Eq.~(\ref{MSH}) reads
\be
M_\mathrm{MSH} = \frac{H^2 R^3}{2G} =\frac{4\pi R^3}{3} \rho 
\ee
where, in the last equality, we used the Friedmann 
equation~(\ref{Friedmann}) in a spatially flat universe. This is 
consistent with the expression of $M_\mathrm{MSH}$ obtained by comparing 
the line element~(\ref{LE}) with the form~(\ref{AbreuVisser}) for general 
spherical geometries.

By comparing the forms~(\ref{GEMmetric}) and (\ref{AbreuVisser}) of the 
line element, one can express the gravitoelectric and gravitomagnetic 
potentials as functions of the Misner-Sharp-Hernandez mass,
\begin{eqnarray}
\Phi &=& \frac{G M_\mathrm{MSH} }{R} \,,\\
&&\nonumber\\
\vec{A} &=& \sqrt{ \frac{2G M_\mathrm{MSH} }{R}} \, \vec{e}_{R} = 
\sqrt{2\Phi} \, \vec{e}_R 
\end{eqnarray}
to first order in the perturbation.

\section{Perturbed FLRW universe}
\label{sec:4}
\setcounter{equation}{0}

We now discuss a toy model of a perturbed FLRW universe, in which there is 
a single, spherically symmetric, scalar perturbation described by the 
post-Newtonian potential $\phi$. The line element in the Newtonian gauge 
is
\be
ds^2 = - \left( 1+2\phi \right)dt^2 +a^2(t) \left( 1-2\phi\right) \left( 
dr^2 +r^2 d\Omega_{(2)}^2 \right) 
\ee
where, for the moment, we allow the spherically symmetric post-Newtonian 
potential to depend on time, $\phi=\phi(t,r)$ with $|\phi| \ll 1$. 
Consistently with the fact that the peculiar velocities of scalar 
perturbations (both primordial dark matter perturbations and 
well-developed galaxies) are usually small in comparison with the Hubble 
flow, the vector perturbations are neglected, which leads to the absence 
of the gravitomagnetic potential $\vec{A}$ in this gauge. This fact is 
consistent with gravitoelectromagnetism when terms of higher order in 
$v/c$ (where $v$ is a typical velocity) are neglected.

The areal radius is 
\be
R(t,r)=a(t) r \sqrt{1-2\phi(t,r)}
\ee
and its gradient
\be
\nabla_{\mu}R = \frac{  \dot{a} r \left( 1-2\phi \right) -ar\dot{\phi} }{ 
\sqrt{1-2\phi}} \, \delta_{0\mu} 
+\frac{a\left( 1-2\phi -r\phi'\right)}{\sqrt{1-2\phi}} \, \delta_{1\mu} 
\ee
(where a prime denotes differentiation with respect to the comoving 
radius $r$ of the FLRW background) gives
\be
\nabla^c R \nabla_c R = 
1-H^2 R^2 \left( 1-2\phi \right) +2HR^2 \dot{\phi} -2r\phi' 
\ee
to first order. Equation~(\ref{MSH}) then gives the 
Misner-Sharp-Hernandez mass
\be
M_\mathrm{MSH} (t,r) = \frac{H^2 R^3}{2G} +\frac{r R\phi'}{G} 
-\frac{H R^3}{G} \left( H \phi +\dot{\phi} \right) 
\,.\label{decomposition}
\ee
The first contribution to the right hand side has cosmological nature (in 
a spatially flat universe, this is the mass of the cosmic fluid enclosed 
by the sphere of radius $R$); the second contribution is purely local, 
while the third contribution is mixed. Thus far, we have performed an 
expansion in powers of $\phi$, keeping only linear terms. We now restrict 
to regions much smaller than the Hubble radius, obtaining two expansions 
with smallness orders ${\cal O}\left( \phi \right)= {\cal O}\left( r\phi' 
\right) $ and $HR$. The mixed term $-\frac{H R^3}{G} \left( H\phi 
+\dot{\phi} \right)$ is of higher order than the two previous terms and is 
usually discarded unless tiny relativistic effects are searched for in 
cosmology \cite{Adamek:2013wja, Adamek:2014xba, Adamek:2015eda, 
Bentivegna:2015flc, Giblin:2015vwq}.

As in any spherically symmetric spacetime, the line element can be 
written\footnote{Since the FLRW geometry is not static, Martel-Poisson and 
Painlev\'e-Gullstrand observers cannot be introduced, but the spherical  
metric can always be cast in the form~(\ref{AbreuVisser}).} 
in the Painlev\'e-Gullstrand form~(\ref{AbreuVisser}) \cite{Abreu:2010ru, 
Nielsen:2005af}, which becomes
\begin{eqnarray}
ds^2 &=& -\left[ 1-H^2R^2 -2r\phi' +2HR^2 \left( H\phi +\dot{\phi} 
\right) \right]  d\bar{t}^2 \nonumber\\
&&\nonumber\\
&\, & \pm 2 \sqrt{ H^2R^2 +2r\phi' 
-2HR^2 \left(  H \phi 
+\dot{\phi} \right) } \, d\bar{t} dR \nonumber\\
&&\nonumber\\
&\, &  + dR^2+ R^2 d\Omega_{(2)}^2 \,. \label{astrocazzo}
\end{eqnarray}
At this stage, we do not yet have gravitoelectromagnetism, which requires 
the metric to be Minkowskian with small corrections. By neglecting the 
time dependence of $H(t)$ and $\phi(t,r)$, one makes the now familiar 
approximations
\be
H(t) \simeq H_0 \,, \quad\quad H_0R \ll 1\,, \quad\quad 
\dot{\phi}\simeq 0 \,, \quad\quad HR\phi \simeq 0 \,,
\ee
obtaining
\begin{eqnarray}
ds^2 & \simeq & ds^2_{(0)} = -\left( 1-H_0^2 R^2 -2r\phi' \right) 
d\bar{t}^2 \nonumber\\
&&\nonumber\\
&\, & \pm 2 \sqrt{ H_0^2R^2 +2r\phi'} \, d\bar{t}dR +dR^2 +R^2 
d\Omega_{(2)}^2  \,.\nonumber\\
&&
\end{eqnarray}
The usual identification of the gravitoelectromagnetic potentials follows:
\begin{eqnarray}
\Phi &=& \frac{H_0^2R^2}{2} +r\phi' \,,\\
&&\nonumber\\
\vec{A} &=& \pm \sqrt{ H_0^2 +\frac{2r\phi'}{R^2} } \, \vec{X} \,.
\end{eqnarray}
Again, the gravitomagnetic potential is purely radial, giving  
gravitomagnetic field $B_{(g)}=\vec{\nabla} \times \vec{A} =0 $. 
If we assume that the FLRW perturbation is due to a single (constant) 
point mass $m$, then $\phi =-Gm/r$ and $\Phi \simeq 
\frac{H_0^2R^2}{2}+\frac{Gm}{R}$. 

The decomposition~(\ref{decomposition}) of the Misner-Sharp-Hernandez mass 
in three contributions was performed in Ref.~\cite{Faraoni:2015kva} in the 
context of the potential problem that $N$-body simulations of large scale 
structures are Newtonian, even though they span volumes larger than the 
Hubble volume at the redshift of structure formation 
\cite{Chisari:2011iq,Green:2011wc, Adamek:2013wja, 
Adamek:2014xba, Adamek:2015eda, 
Bentivegna:2015flc, Giblin:2015vwq}. There, a ``potential'' $\sim 
\frac{H^2 R^2}{2} 
+\frac{Gm}{R}$ was introduced {\em ad hoc} to quantify the degree of 
``non-Newtonianity'' of dark matter perturbations (the result was that the 
Newtonian simulations of large scale structures are adequate) 
\cite{Faraoni:2015kva}. It was not realized, however, that this fictious 
potential appears in the gravitoelectromagnetic description of cosmology 
in the approximation in which the FLRW background is replaced with a de 
Sitter one.

\section{Conclusions}
\label{sec:5}
\setcounter{equation}{0}

We have examined FLRW cosmology from the perspective of the most 
well-known version of gravitoelectromagnetism in linearized GR. The 
alternative formulation of gravitolectromagnetism using electric and 
magnetic parts of the Weyl tensor with respect to a given observer ({\em 
e.g.}, \cite{Maartens:1997fg}) does not apply to FLRW universes, in which 
the Weyl tensor vanishes identically \cite{Waldbook}.

In retrospect, even the ``standard'' picture of gravitoelectromagnetism is 
not so standard when applied to FLRW universes. In fact, one must replace 
the exact FLRW manifold with its instantaneous de Sitter approximation, 
which implies that the matter stress-energy tensor must necessarily be the 
effective one associated with a cosmological constant $\Lambda=3H_0^2$, 
and not that of a dust. Moreover, in order for the spacetime metric to be 
the Minkowski one plus small perturbations, one must restrict oneself to 
spacetime regions small in comparison with the Hubble radius $H^{-1}$, 
instead of large regions far away from localized energy distributions.

A freely falling (geodesic) observer will always see the spacetime metric 
as the flat one plus small perturbations in a local expansion 
\cite{MTW,Marzlin:1994ia}. Freely-falling observers are determined 
up to a Lorentz boost ({\em e.g.}, \cite{Weinberg}). In FLRW universes, it 
is natural to consider freely falling radial observers, to which are 
associated special coordinates in cosmology 
\cite{Gautreau1,Gautreau2,Parikh:2002qh, 
Faraoni:2020ehi,Grib:2020kzh,Vachon:2021bya}. Since the FLRW universe is 
approximated locally with an osculating de Sitter space, which is locally 
static, one can introduce Martel-Poisson observers and their special 
subclass, the Painlev\'e-Gullstrand observers \cite{Faraoni:2020ehi}. It 
is rather natural to formulate gravitoelectromagnetism in the 
Painlev\'e-Gullstrand gauge. This is different from the usual Lorentz 
gauge and is more similar to the Bakopolous-Kanti gauge 
\cite{Bakopoulos:2014exa,Bakopoulos:2016rkl}. In asymptotically flat 
linearized GR, the Bakopoulos-Kanti gauge is valid only {\em in vacuo} but 
the situation is different in cosmology, in which the metric components 
have two different orders of smallness.

As expected from spatial isotropy, the gravitoelectric field is purely 
radial and the gravitomagnetic field vanishes identically as a 
consequence of the gravitomagnetic potential $\vec{A}$ being radial. Due 
to the spherical symmetry of FLRW spaces about every spatial point, one 
can introduce the Misner-Sharp-Hernandez quasilocal mass \cite{MSH1,MSH2} 
and we have expressed the gravitoelectromagnetic potentials $\Phi, 
\vec{A}$ in 
terms of it.

It is also interesting to consider perturbed FLRW universes from the 
perspective of gravitoelectromagnetism. For 
simplicity, we have considered the  situation of a single 
spherically symmetric metric perturbation. The analysis of 
Ref.~\cite{Faraoni:2015kva} 
of the physics of $N$-body simulations, which are Newtonian even though 
the box used is a few Hubble scales in size, was based on the splitting of 
the Misner-Sharp-Hernandez mass into local and cosmological perturbations, 
discarding a much smaller contribution \cite{Faraoni:2015kva}. Here the 
fictitious potential used in \cite{Faraoni:2015kva} has been shown to 
coincide with the 
gravitoelectrostatic potential of FLRW universes, making more meaningful 
the discussion of \cite{Faraoni:2015kva}. One could generalize the 
discussion to arbitrary (small) cosmological perturbations, in which case 
the Misner-Sharp-Hernandez mass (defined only in spherical symmetry) 
cannot be used. However, one can use its Hawking-Hayward quasilocal 
generalization \cite{Hawking, Hayward:1993ph, Hayward:1994bu}, as done in 
Ref.~\cite{Faraoni:2015kva}. We do not repeat the discussion of 
\cite{Faraoni:2015kva} here, the conclusion being the rather obvious 
generalization of the gravitoelectromagnetic potentials to the 
non-spherical 
case.

To conclude, even though gravitoelectromagnetism in FLRW cosmology could 
be expected to be rather trivial, it is not: we have uncovered several 
non-trivial aspects and many differences with respect to the usual 
discussion of linearized GR in asymptotically flat spaces.

\begin{acknowledgments} 

This work is supported, in part, by the Natural Sciences \& Engineering 
Research Council of Canada (grant no. 2016-03803 to V.F.) and by a 
Bishop's University Graduate Entrance Scholarship to S.J.

\end{acknowledgments}



\end{document}